\documentclass{aa}
\usepackage{graphicx}
\newcommand{\be}{\begin{equation}}
\newcommand{\ee}{\end{equation}}

\begin{document}
\thesaurus{12(02.01.2;02.13.2;08.06.2;09.10.1;11.14.1;11.10.1)}

\title{Magnetized accretion-ejection structures}
\subtitle{V. Effects of entropy generation inside the disc}

\author{Fabien Casse \and Jonathan Ferreira}
\offprints{Fabien.Casse@obs.ujf-grenoble.fr}

\institute{Laboratoire d'Astrophysique de l'Observatoire de Grenoble BP53,
  F-38041 Grenoble cedex 9, France}

\date{Received 28 April 2000/Accepted 26 July 2000}
\titlerunning{Entropy generation in MAES}
\authorrunning{F. Casse \& J. Ferreira}
\maketitle

\begin{abstract}

In this paper, steady-state MHD calculations of non-relativistic magnetized
accretion discs driving jets are presented. For the first time, an energy
equation describing the effects of entropy generation along streamlines is
included. Using a simplified approach, we showed that thermal effects have a
tremendous influence on the disc capability to feed jets with mass. 

The disc ejection efficiency is measured by the parameter $\xi= d \ln
\dot{M}_{a}/d \ln{r}$, where $ \dot{M}_{a}(r)$ is the local disc accretion
rate. While previous self-similar solutions were only able to produce jets
with $\xi \sim 0.01$, solutions with a coronal heating display huge
efficiencies up to $\xi \sim 0.5$. Moreover, taking thermal effects into
account allows to obtain both fast and slow magnetic rotators.

Since most of the jet properties (like asymptotic velocity or degree of
collimation) depend on the mass load, it arises from this study that any
quantitative result requires a detailled analysis of the disc energetics.

\end{abstract}

\section{Introduction}

\subsection{Jets and discs}

Astrophysical jets are observed in a wide variety of objects, from young
protostars surrounded by a circumstellar accretion disc to compact objects
in either, close binary systems or active galactic nuclei. In all these
systems, several observational evidences point towards strong links between
accretion and ejection phenomena (Hartigan et al. \cite{Har}, Mirabel et
al.  \cite{Mir} and Serjeant et al. \cite{Sarj}). Thus, theorists must
provide models where accretion and ejection can be simultaneously explained
in a self-consistent way. Precisely, these models must provide a clear
picture of how a fraction of the disc matter becomes an ejected flow, but
also of the energetics (eg, the ratio of the jet kinetic power to the
accretion power). Indeed, some problematic objects show powerful jets
produced from discs showing a much lower activity than expected (Rutten et
al. \cite{rutten}, Di Matteo et al. \cite{DiMat}). We will call this the
energy problem.  

It is well known that the ``standard'' accretion disc model of Shakura \&
Sunyaev (\cite{Shak}) cannot explain jets. In this model, all the mass is
accreted towards the central object while the liberated mechanical power
(called accretion power) is radiated at its surfaces. However, it as been
recognized that, under certain circumstances, the disc could become
geometrically thick, hence allowing for an efficient advection of the
thermal energy (Abramowicz et al. \cite{Abra}). This has an important
observational consequence. Indeed, these advection-dominated accretion flow
(ADAF) models allow a sizeable fraction of the accretion power to be actually
advected onto the central object. Thus, such a power would be
observationaly ``lost'' if this object happens to be a black
hole. Moreover, it has been argued that such ADAFs would be hot enough to,
hopefully, provide a launching mechanism for jets (Narayan \& Yi
\cite{NarI}). As a consequence, one might have jets with a high kinetic power 
arising from low-luminosity discs. All this reasoning is fine, but no
comprehensive work has been done showing such a possibility. There is
however an alternative to this energy problem. 

\subsection{Models of magnetized discs driving jets}

First, we recall that a very efficient launching mechanism is the presence
of a large scale bipolar magnetic field threading the accretion
disc (for an alternative view, see Lery et al. \cite{Lery}). This magnetic
field can be generated by a dynamo effect (see Rekowski 
et al. \cite{Rekowski}) and/or by advection of interstellar magnetic
field (Mouschovias \cite{Mous}). Indeed, since the seminal work of Blandford \& Payne (\cite{Bland}), it has been showed that, combined with the disc
rotation, such a field could {\bf (1)} extract the disc angular momentum,
{\bf (2)} provide an acceleration force to a fraction of the mass and {\bf
(3)} maintain the jets self-collimated by a hoop-stress. Thus, explaining
collimated jets would require to revisit the standard model by
self-consistently incorporating magnetic fields. We call these magnetized
objects, magnetized accretion-ejection structures (MAES).

The first attempt to describe such objects was made by K\"onigl (\cite{Ko})
and followed by Wardle \& K\"onigl (\cite{Ward}), Li (\cite{LiI},
\cite{LiII}). However, these works had two drawbacks. First, they
did not self-consistently treat the disc vertical equilibrium, thereby
obtaining huge parameter spaces and, second, the overall energy budget was
not considered.

A complete description of MAES has been undergone in a series of papers
where, thanks to a self-similar ansatz, all dynamical terms were included
in the magnetohydrodynamic (MHD) equations (Ferreira \& Pelletier
\cite{FP93a}, \cite{FP95}, Ferreira \cite{F97}, Casse \&
Ferreira \cite{Fab}). In particular, it has been shown that a necessary
condition for jet production is the reversal of the sign of the magnetic
torque at the disc surface: negative
at the disc midplane (angular momentum extraction), it must become
positive at the surface in order to magnetically propell matter. This
condition provides a strong constraint on the properties of the MHD
turbulence believed to exist inside discs. Casse \& Ferreira (\cite{Fab})
provided a comprehensive scan of the parameter space of ``cold'' MAES,
i.e. where enthalpy plays no role in jet dynamics. They assumed adiabatic
magnetic surfaces and showed that, depending on the MHD turbulence
properties, two kinds of ``cold'' MAES could exist: 
\begin{itemize}
\item[(1)] discs where the magnetic torque due to the jets is much larger
  than the local ``viscous'' torque (of turbulent origin). These MAES are
  called ``non-dissipative''. Indeed, almost all the available power feeds
  the jets and only a tiny fraction is locally dissipated and radiated by
  the disc (see also Ferreira \& Pelletier \cite{FP93a} and
  Ferreira \cite{F97}). Another important consequence of the larger
  magnetic torque is a disc density smaller than in a standard disc (for
  the same accretion rate). A non-dissipative MAES could then be optically
  thin, and is thus expected to provide a spectrum very much comparable to
  that of an ADAF.  
  
\item[(2)] discs where the magnetic torque is comparable to the viscous
  torque, called ``dissipative'' MAES. Such structures dissipate as
  radiation a power comparable to that carried away by the jets. No
  solution has been found with a dominant viscous torque.
\end{itemize}

Thus, MAES quite straightforwardly address the energy pro\-blem mentionned 
earlier. Instead of advecting energy onto the central object like ADAF do,
MAES put a sizeable fraction of the accretion energy into jets. In fact,
everything depends on the properties of the MHD turbulence inside the disc,
which are still unknown.

\subsection{The ``cold jet'' approximation}

Nonetheless, all these results (as well as all previous works on disc winds)
rely on what happens to be a very important assumption, namely the ``cold
jet '' approximation. Usually, in the literature, a ``cold jet'' is a jet
where enthalpy plays no role in its launching (ie, is negligible in the
Bernoulli equation). Thus, in order to get ejection from the disc surface,
the opening angle, defined as the angle between the poloidal magnetic field
and the vertical axis, must be bigger than 30$^o$ (Blandford \& Payne
\cite{Bland}). However, the crucial question is the amount of mass that
can be loaded in the jet. This is {\bf never} addressed in any steady-state
numerical simulation, since the underlying accretion disc is merely a
boundary condition. This mass load critically depends on the force balance
at the disc surface, where matter can still cross the magnetic field
lines. In this transition layer from a resistive disc to an ideal MHD 
jet, the only force allowing matter to go up is the plasma pressure
gradient (Ferreira \& Pelletier \cite{FP95}). Thus, the vertical profile of
the temperature does play a role in the launching process. Since it mainly
occurs in the resistive MHD region, a negligible enthalpy in the ideal MHD
zone does not mean that it had no influence on jet formation. 

In this paper, we will show that the precise treatment of the energy
equation must be done correctly, especially at the disc surface. This is a
{\it sine qua non} condition in order to get realistic quantitative answers
concerning the capability of accretion discs to launch jets. In particular,
the presence of a hot corona could drastically change the whole
picture. Because cooling and heating processes depend on density and
temperature, the conditions at the disc surface vary with the radius and
depend also on the astrophysical context studied (e.g around a compact
object or a young star). Here, our purpose is just to illustrate this
stinging but crucial point by using some crude prescription. We will
investigate specific astrophysical contexts in forthcoming papers.

In section 2, we recall the basic stationary MHD equations describing a
MAES and define the relevant disc and jet parameters. Section 3 is the
cornerstone of this paper. We introduce the local disc energy equation with
the prescription we used and discuss the constraints imposed by the global
energy conservation. Thermal effects on stationary jet production are then
analyzed. In section 4, we show some typical self-similar MAES solutions
and display two extreme cases. We finally conclude in section 5.

\section{Magnetized accretion-ejection structures}

\subsection{Basic MHD stationary equations}

Thanks to axisymmetry, the vectorial quantities expressed in cylindrical
coordinates ($r$, $\phi$, $z$), can be decomposed into poloi\-dal and
toroidal components, e.g. $\vec{u}=\vec{u_p} + \Omega r \vec{e_{\phi}}$ and
$\vec{B}=\vec{B_p} + B_{\phi} \vec{e_{\phi}}$. A bipolar configuration
allows us to describe the poloidal magnetic field as
\begin{equation}
{\vec B_p} = \frac{1}{r}\vec{ \nabla} a \times {\vec e_{\phi}}  \ ,
\end{equation}
\noindent where $a (r,z)$ is an even function of $z$ and $a=constant$
describes a surface of constant magnetic flux ($a=rA_{\phi}$, $A_{\phi}$
being the toroidal component of the potential vector). The following set of
equations describes a steady-state, non-relativistic MAES, 
\begin{itemize}
\item{Mass conservation 
    \begin{equation} 
    \vec{ \nabla} \cdot \rho{\vec u}= 0
    \label{1}
    \end{equation}}
\item{Momentum conservation 
    \begin{equation} 
    \rho{\vec u}\cdot \vec{ \nabla}{\vec u}=-\vec{ \nabla} P-\rho\vec{
      \nabla}\Phi_G + \vec{J}\times\vec{B} + \vec{ \nabla}\cdot \mathsf{T} 
    \label{2}
    \end{equation}}
\item{Ohm's law and toroidal field induction
    \begin{eqnarray}
      \eta_mJ_{\phi} {\vec e_{\phi}} &=&\vec{ u_p}\times\vec{B_p}\label{3}\\
      \vec{ \nabla}\cdot (\frac{\nu'_m}{r^2}\vec{ \nabla} rB_{\phi}) & = &
      \vec{ \nabla}\cdot\frac{1}{r}(B_{\phi}\vec{ u_p} - \vec {B_p} \Omega
      r) \label{4} 
    \end{eqnarray}}
\end{itemize}
\noindent where $\rho$ is the density of matter, $P$ the thermal pressure,
$\Phi_G=-GM_*/(r^2+z^2)^{1/2}$ the gravitational potential, $\vec J= \vec{
\nabla}\times \vec B/\mu_o$ the current, $\mathsf{T}$ the ``viscous''
stress tensor, $\nu_m= \eta_m/\mu_o$ and
$\nu'_m$ the (anomalous) poloidal and toroidal magnetic diffusivities. All
transport coefficients appearing in the above set of equations, namely
$\nu_m$, $\nu'_m$ and $\nu_v$ (``viscosity'', contained in $\mathsf{T}$),
are assumed to be of turbulent origin. The state equation used is the
perfect gas law
\begin{equation} 
P = nk_B T
\label{5}
\end{equation}
\noindent where $k_B$ is the Boltzmann constant, $n=\rho/m_p$ ($m_p$ being
the proton mass) and $T$ the plasma temperature.

The last equation is the energy equation. Until now, it has been replaced
by a simple prescription on the temperature, na\-mely isothermal or adiabatic
magnetic surfaces (see eg Ferreira \& Pelletier \cite{FP95} for a
discussion on this subject). The aim of the present paper is to study the
thermal effects on jet production. Therefore, we must consider the energy
equation, which can be written as
\begin{equation}
\rho T\frac{dS}{dt} =  \rho T \vec{u_p} \cdot \vec{ \nabla} S = Q
\label{6}
\end{equation}  
\noindent where $S$ is the specific entropy and $Q$ is the local source of
entropy, arising from the difference between heating and cooling processes
(see Sect.~3 for more details).

\subsection{Dynamical parameters of MAES}

The set of equations (\ref{1}) to (\ref{6}) completely describes a
steady-state, non-relativistic MAES. We recall here the definitions of the 
parameters used:
\begin{itemize}
\item{The disc aspect ratio
\be 
\varepsilon= \frac{h(r)}{r}
\ee}
\item{The MHD turbulence level (measuring the strength of the poloidal
    magnetic diffusivity at the disc midplane)
\be
\alpha_m= \left.\frac{\nu_m}{V_Ah}\right|_{z=0}
\ee}
\item{The magnetic Prandtl number (measuring the strength of the turbulent
    viscosity) 
\be
{\cal P}_m = \left.\frac{\nu_v}{\nu_m}\right|_{z=0}
\ee}
\item{The magnetic diffusivity anisotropy
\be
\chi_m=\left.\frac{\nu_m}{\nu'_m}\right|_{z=0}
\ee}
\end{itemize}
\noindent where $h(r)$ is the vertical disc scale height and $V_A$ the
Alfv\'en speed at the disc midplane. These four parameters are defined
locally and could, in principle, vary along the disc. In our study however,
they are assumed constant. Also, while the disc aspect ratio $\varepsilon$ 
remains free (because we did not treat the real energy equation), the three
other turbulence parameters must remain free, until the development of a full
theory of MHD turbulence. All other disc and jet quantities can be
expressed in terms of these parameters (see Casse \& Ferreira
\cite{Fab}, hereafter CF). As an example, the usual alpha coefficient of
Shakura \& Sunyaev is given by
\be
\alpha_v = \left. \frac{\nu_v}{C_s h} \right|_{z=0} = {\cal P}_m \alpha_m^*  
\ee
\noindent where $\alpha_m^* = \nu_m/C_s h = \alpha_m (\mu/\gamma)^{1/2}$, 
$\mu= (V_A/\Omega_K h)^2$ is a measure of the magnetic field strength
($\Omega_K$ is the Keplerian rotation rate) and $\gamma$ the adiabatic
index. Note that, since steady state MAES require fields close to
equipartition (ie. $\mu \leq 1$, Ferreira \& Pelletier \cite{FP95}), one
has $\alpha_m^* < \alpha_m$.

\subsection{Generalized ideal MHD integrals}

Inside the disc, accreting matter must cross the field lines. Thus, a
stationary configuration requires turbulent magnetic diffusivities. This
turbulence is believed to be triggered inside the disc by some MHD
instability, but decays vertically on (say) a disc scale height. In the
ideal MHD regime characterizing the jets, integrals of motion appear in the
basic MHD equations. These are 
\begin{eqnarray}  
\vec{u_p} &=& \frac{\eta (a)}{\mu_o\rho}\vec{B_p}\label{up}\\
\Omega_*(a) &=& \Omega -\eta\frac{B_{\phi}}{\mu_o\rho r} \\ 
\Omega_*r^2_A &=& \Omega r^2 - \frac{rB_{\phi}}{\eta}
\end{eqnarray} 
\noindent where $\eta(a)=\sqrt{\mu_o\rho_A}$ is a constant along a magnetic
surface and $\rho_A$ is the density at the Alfv\'en point, where the
poloidal velocity $u_p$ reaches the poloidal Alfv\'en velocity
$V_{A,p}$. The rotation rate of a magnetic surface is $\Omega_*(a)$ and
$r_A$ is the Alfv\'en radius. Following Blandford \& Payne (\cite{Bland}),
we introduce the following jet parameters
\begin{eqnarray}
&\lambda & = \frac{\Omega_*r^2_A}{\Omega_or^2_o}\nonumber\\ 
&\kappa & = \eta\frac{\Omega_or_o}{B_o}\\ 
&\omega_A& = \frac{\Omega_*r_A}{V_{Ap,A}}\nonumber \ .
\end{eqnarray}
\noindent The subscript ``o'' refers to quantities evaluated at the disc
midplane, ``A'' at the Alfv\'en surface. These parameters are directly
related to the disc parameters (see CF). The last ideal MHD invariant is
the Bernoulli integral. This invariant is modified by the presence of
heating, contrary to the three former invariants. By projecting the
momentum conservation on a magnetic surface and using thermodynamics
relations, one gets (Chan \& Henriksen \cite{Chan}, Sauty \& Tsinganos 1994)
\begin{equation} 
E(a) = \frac{u^2}{2} + \frac{\gamma}{\gamma -1}\frac{P}{\rho} + \Phi_G
-\Omega_*\frac{rB_{\phi}}{\eta} - {\cal H}(s,a)    
\label{Bernou}
\end{equation}
\noindent where the enthalpy (second term) will be hereafter represented by
$H$ and
\be
{\cal H}(s,a)= \int^s_{s^+} \frac{Q(s',a)}{\rho(s',a)u_p(s',a)}ds'
\ee
\noindent is a heating term that depends on the curvilinear coordinate $s$
along a given magnetic surface (labelled by a constant $a(r,z)$). Here, $s^+$
represents the coordinate of the transition between the resistive and the
ideal MHD regimes (above the disc surface). Jets from quasi-Keplerian
accretion discs are energetically possible if 
\begin{equation}
E(a) = \Omega_o^2r_o^2\left(\lambda - \frac{3}{2} + \frac{\gamma}
{\gamma -1}\frac{T^{+}}{T_{o}} \varepsilon^2 \right) \ . 
\end{equation}
\noindent is positive at the base of the jet. Thus, either the magnetic
lever arm $\lambda$ is larger than $3/2$ or some heating occurred in the
underlying resistive layers. This last possibility arises whenever the disc
is already hot (ie. thick $\varepsilon \sim 1$ or slim $\varepsilon < 0.3$
discs) and/or if some coronal heating leads to an increase of the
temperature $T^+$ at the disc surface.

\section{Entropy generation in MAES}

\subsection{Local entropy generation}

The local energy equation (\ref{6}) is
\begin{eqnarray}
  Q & = &\rho T \frac{d S}{d t}=\rho\frac{d H}{dt} - \frac{dP}{dt}
  \nonumber \\
  & = &  \frac{\gamma}{\gamma -1}\frac{k_B}{m_p}\rho\vec{u_p}\cdot \vec{
  \nabla}T - \vec{u_p}\cdot\vec{ \nabla} P \ .
  \label{equi}
\end{eqnarray} 
\noindent In a thin (or slim) accretion disc, the gradient of any plasma
quantity is mostly vertical. So the main influence of an entropy source
$Q$ occurs whenever the plasma velocity becomes vertical namely, at the
disc surface. There, a non-vanishing $Q$ will have two distinct and
important consequences. First, it raises the plasma temperature ($T^+$)
and, second, it causes an increase of the vertical plasma pressure 
gradient. Now, the vertical component of the momentum conservation writes
\begin{equation}
\rho u_z \frac{\partial u_z}{\partial z} \simeq -\frac{\partial P}{\partial
  z} - \rho\Omega_K^2 z - \frac{\partial}{\partial
  z}\frac{B^2_{\phi}+B^2_{r}}{2 \mu_o} \ .
\end{equation}
\noindent In this equation, the only force that can gently expell matter
from the disc is the vertical plasma pressure gradient. Thus, increasing
this gradient through an entropy source $Q$ allows new configurations to
exist. For example, with parameters ($\varepsilon$, $\alpha_m$, 
${\cal P}_m$, $\chi_m$) similar to those of a ``cold'' MAES (ie. $Q\simeq
0$), a solution with an entropy source would display a larger vertical force,
allowing thereby a larger mass flux $\kappa$. Or, with the same entropy
source, one could get a steady ejection from a MAES with a larger magnetic
pinching, corresponding to turbulence parameters impossible to achieve with
$Q\simeq 0$. 

Thus, we see that both the amplitude and the vertical profile of the
entropy source $Q$ are important features for stationary MAES. In fact,
there is a local entropy source $Q$ only if there is a slight discrepancy
between the local heating and cooling terms, namely 
\be  
Q = (\Gamma_{eff} + \Gamma_{turb} + \Gamma_{ext}) - (\Lambda_{rad} +
\Lambda_{turb}) 
\label{madaf}
\ee
\noindent where 
\be 
\Gamma_{eff} = \eta_mJ^2_{\phi} + \eta'_mJ^2_p + \eta_v
\left|r\vec{ \nabla}\Omega\right|^2  
\label{gammaeff}
\ee
\noindent is the effective Joule and viscous heating, $\Gamma_{turb}$ is a
turbulent heating term which cannot be described by simple anomalous
transport coefficients (eg, the term $\Gamma_{eff}$), $\Gamma_{ext}$ is an
external source of energy (typically due to some illumination by UV or
X-rays, hence maximum at the disc surface),  
\be
\Lambda_{rad}= \vec{ \nabla} \cdot \vec S_{rad}
\ee
\noindent is the radiative cooling ($\vec S_{rad}$ being the radiative flux)
and $\Lambda_{turb}$ is a cooling due to a turbulent energy transport,
which is most probably also taking place inside turbulent discs (see
eg. Sha\-kura, Sunyaev \& Zilitinkevitch \cite{Shak78}).  

If the disc is optically thick, the radiative flux $\vec S_{rad}$  can be
written in a simple way using the diffusion approximation (eg. Ferreira
\& Pelletier \cite{FP93a}). However, both the amplitude and the vertical
profile of the turbulent energy transport $\Lambda_{turb}$ are
unknown. This is really problematic, since this term could be the leading
cooling term (like convection, see Ferreira \& Pelletier \cite{FP95}). On
the other hand, magnetized discs may well have active coronae, hence a non
negligible $\Gamma_{turb}$ mainly acting at the disc surface (Galeev et
al. \cite{Galeev}, Heyvaerts \& Priest \cite{HeyPriest}, Miller \& Stone
\cite{Miller}). Therefore, even if we neglect any external influence, the
only term whose amplitude and vertical profile are really known is
$\Gamma_{eff}$. Thus, following Narayan \& Yi (\cite{NarI}) for ADAFs, we
define a parameter $f$ as  
\be
f = \frac{\int_V Q dV} {\int_V \Gamma_{eff} dV} = \frac{ \int_V Q dV}
{P_{diss}}  
\ee
\noindent where the integral is made over the volume $V$ occupied by the
disc and $P_{diss}$ is the total power which could be dissipated inside the
disc. With this prescription, the amplitude of the entropy source $Q$ is
controlled by $f$, but not its vertical profile. This profile depends on
the relative importance of the various terms appearing in Eq.~(\ref{madaf})
and, as such, is beyond our actual knowledge. To tackle this difficult
problem, we will assume the profile and discuss the corresponding physical
implications on jet production.

\subsection{Global energy conservation}

A strong constraint on this parameter $f$ arises from the global energy
conservation. If we integrate the local energy conservation equation over
the whole volume $V$, we get the following energy budget 
\be
P_{acc} \ +\  P_{ext} = 2 P_{jet} \ + \  2P_{rad} 
\label{budget}
\ee
\noindent where $P_{acc}$ is the power liberated through the accretion
process, 
\be
P_{ext} = \int_V \Gamma_{ext} dV
\label{ext}
\ee
\noindent is the total input of energy due to an external source, $P_{jet}$
is the total power (mechanical, magnetic and thermal) carried away by one
jet (hence leaving the disc at its surface) and $P_{rad}$ the luminosity
radiated at one surface (see Appendix A). 

In this paper, we do not consider any external source of energy, thus
$P_{ext} = 0$. Moreover, we do not expect in a quasi-Keplerian flow the
accreted turbulent energy to be significant with respect to the mechanical
energy. Thus, we assume that the total turbulent power is negligible,
namely 
\be
P_{turb}= \int_V (\Gamma_{turb} - \Lambda_{turb}) dV = 0 \ .
\label{turb}
\ee

The characteristic power, liberated by a MAES located a\-round a
central mass $M_*$, between the radii $r_i$ and $r_e$ and fed with an
accretion rate  $\dot M_{ae}=\dot M_a(r_e)$, is 
\be
P_{lib} \equiv  \eta_{lib}\frac{GM_*\dot{M}_{ae}}{2 r_i}       
\ee
\noindent where the efficiency $\eta_{lib}$ is given in Appendix B. In
discs producing jets, the accretion rate is a function of the radius. We
therefore define the local ejection efficiency as
\be
\xi= \frac{d\ln \dot{M}_{a}}{ d\ln{r}} \ .
\ee
\noindent In terms of the fiducial value $P_{lib}$, the global energy
conservation (\ref{budget}) of thin (or even slim) MAES writes (see
Appendix B for details)  
\be
\frac{P_{acc}}{P_{lib}} = (1 - \xi)\left (1 + \frac{1}{2}\frac{\varepsilon
  \Lambda}{1+\Lambda} \right )
\ee
\noindent providing jets powered with
\be
\frac{2P_{jet}}{P_{lib}} = \frac{\Lambda}{1+\Lambda}\left|
  \frac{B^+_{\phi}}{qB_o}\right|\  +\  \frac{2\gamma}{\gamma-1}
\frac{T^+}{T_o}\xi\varepsilon^2 \ -\  \xi 
\label{jetpower}
\ee
\noindent and a disc luminosity of 
\be
\frac{2P_{rad}}{P_{lib}} = (1 - f) \frac{P_{diss}}{P_{lib}} =
\frac{P_{acc}-2P_{jet}}{P_{lib}} 
\label{lum}
\ee
\noindent Here, the parameter $\Lambda$ is the ratio of the magnetic torque
to the viscous torque at the disc midplane and $q$ is a measure of the
radial current density at the disc midplane (see CF for more details). 

From the above energy budget, it becomes obvious that two extreme cases can
be accounted for, namely
\begin{itemize}
\item $f \ll 1$: cooling and heating processes balance each other very
  effectively at every point, so that there is no entropy generation at all
  ($Q\simeq 0$). In this case, the dissipated power $P_{diss}$ is finally
  radiated at the disc surfaces. 
\item $f = 1$: a large entropy source is generated in this case, because of
  a turbulent transport more efficient than radiative cooling and the
  presence of another heating process (eg. coronal heating). As a
  consequence, no power is radiated at all, the accretion power being
  transferred into the jets.
\end{itemize}
The case $f>1$ can still be obtained but requires an extra source of
energy ($P_{ext}$ or $P_{turb}$). This is also true when $f<1$ but the disc
luminosity exceeds that provided by Eq.~(\ref{lum}).

\subsection{Constraints on steady-state jet production}

The trans-Alfv\'enic condition is modified by the possible presence of a
heating term along the flow. This can be directly seen in the function $g$,
defined as $\Omega = \Omega_*(1-g)$ and evaluated at the Alfv\'en surface
\begin{eqnarray}
  g_A^2 &=& 1 - \frac{3}{\lambda} - \frac{1}{\omega^2_A} +
  \frac{2}{\lambda^{3/2}(1+z^2_A/r^2_A)^{1/2}}  \nonumber \\
  & & + \frac{2}{\lambda} \left [ \frac{\gamma}{\gamma -1} 
    \frac{T^+ - T_A}{T_o} \varepsilon^2 
    + \frac{ {\cal H}(s_A,a)}{\Omega_o^2r_o^2} \right ]
  \label{ga}
\end{eqnarray}
\noindent The subscripts ``o'' and ``A'' denote a quantity measured,
respectively, at the disc midplane and at the Alfv\'en surface. A necessary
condition for trans-Alfv\'enic cold jets is usually expressed as 
\be
\omega^2_A > 1 \ , 
\ee
\noindent implying these jets are fast rotators (Michel \cite{Michel},
Pelletier \& Pudritz \cite{PP92}, Ferreira \cite{F97}). However, this
condition is only strictly valid for very large magnetic lever arms
($\lambda \rightarrow \infty$). The generalized trans-Alfv\'enic condition
is in fact  
\be
g_A^2 > 0
\ee
Thus, Eq.~(\ref{ga}) shows that entropy generation inside the disc (leading
to a large $T^+$) or alternatively, some coronal heating (significant ${\cal 
  H}(s_A,a)$), allows the fulfillment of the above condition with $\omega_A
< 1$. Therefore, we expect slow rotators to emerge as
(magneto-)thermally driven jets. 

Following the same reasoning than Ferreira (\cite{F97}) in Sect.~3, but
considering the generalized Bernoulli equation, we obtain an inequality
constraining the magnetic lever arm 
\begin{eqnarray}
  \frac{\lambda}{\lambda - 1} &<& \lambda\  -\ 3\ +\ \frac{2}{\lambda^{1/2}
    (1+z^2_A/r^2_A)^{1/2}} \nonumber\\
& & +\  \frac{2\gamma}{\gamma-1}\frac{T^+}{T_o}\varepsilon^2\ +\ 2\frac{
    {\cal H}(s_A,a)}{\Omega_o^2r_o^2}\ .   
\end{eqnarray}
This quite complex relation is equivalent to $\lambda > \lambda_{min}= 1
+ \zeta$. This shows that, to produce super-Alfv\'enic jets, there must
be a minimum magnetic lever arm $\lambda_{min}$. Namely, steady-state MAES
cannot display an ejection efficiency larger than
\be
\xi_{max}= \frac{1}{2\zeta} \frac{\Lambda}{1 + \Lambda}
\ee
The exact value of $\zeta$ is strongly dependent on $Q$ (both amplitude $f$
and vertical profile). Basically, the more heat is added and the more mass
is ejected. For example, for $T^+ \leq T_o$, $\zeta= 3$ while for $T^+ \simeq
\varepsilon^{-2} T_o$, $\zeta\simeq 0.3$. Thus, as already pointed out, the
major constraints on $\xi$ arise from the disc vertical equilibrium in the
resistive MHD regime and the energy balance.

\section{Numerical results}

\begin{figure}[t]
   \includegraphics[angle=-90,width=\columnwidth]{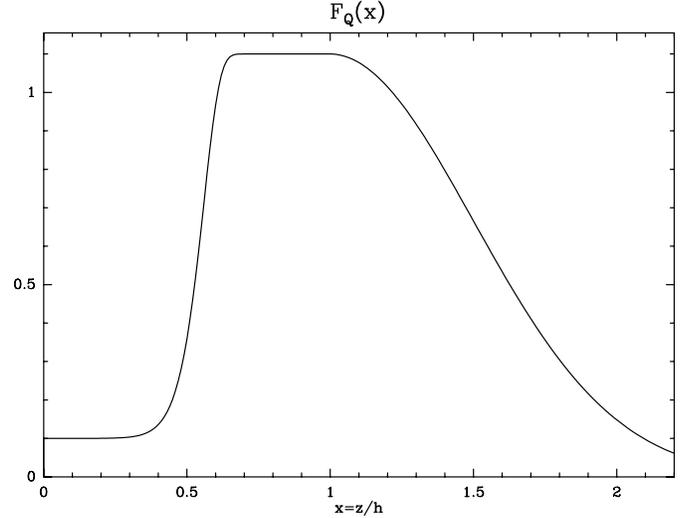}
   \caption{Vertical profile of the entropy generation function $Q$ for a
     given ejection index $\xi=0.1$. The value of this function at the disc
     midplane ($x=0$) is imposed by the self-similar radial scalings (see
     text). The integral of this function is measured by the parameter
     $f$. The particular shape shown here (increasing towards the disc
     surface) mimics a coronal heating.}
   \label{Q}
\end{figure}

\begin{figure}[t]
   \includegraphics[angle=-90,width=\columnwidth]{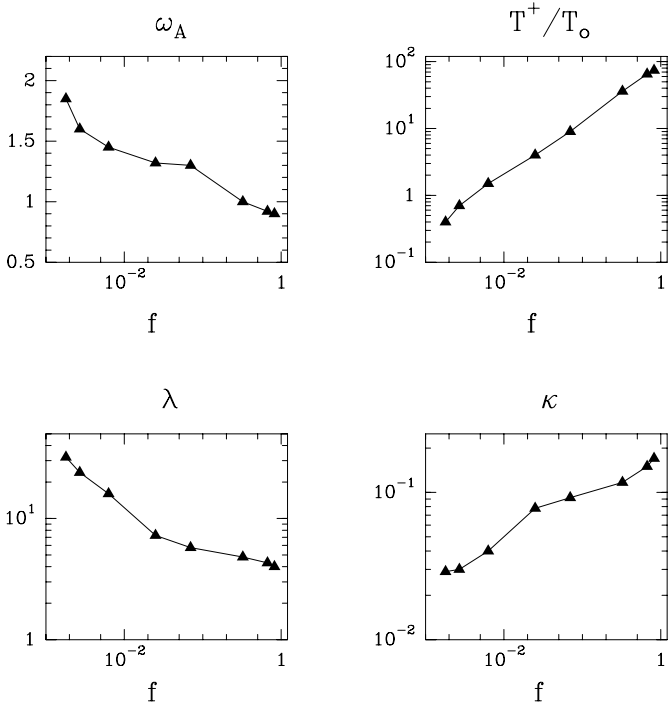} 
   \caption{Variation of several quantities with the entropy parameter $f$
     for a given magnetic configuration: $\varepsilon=0.1$, $\alpha_m=1.5$
     ($\alpha_m^*=0.8$), ${\cal P}_m=1$, $\chi_m=1.5$. Non-relativistic
     jets from Keplerian accretion discs can be either fast rotators
     ($\omega_A > 1$) or slow rotators ($\omega_A < 1$), depending on the
     amplitude of $f$. The coronal temperature increases
     monotonously with $f$. The plots in mass load $\kappa$ and magnetic
     lever arm $\lambda$ illustrate the fact that the more entropy is
     generated the more mass is loaded (see Sect.~3.1).}
   \label{Evo}
\end{figure}

\begin{figure}[t]
  \resizebox{\hsize}{!}{\includegraphics[angle=-90]{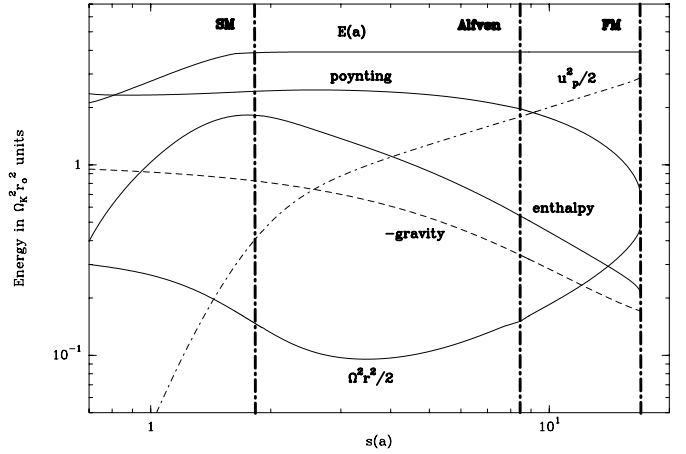}}
  \caption{Bernoulli invariant $E(a)$ and its components along a given
    magnetic surface ($s(a)=z(a)/h_o$), for the hottest jet ($f=0.82$) in
    Fig.~(\ref{Evo}). The quantities are normalized to the square of the
    angular velocity at the footpoint of the magnetic surface ($\Omega_o^2
    r_o^2$). Enthalpy is of the same order than the MHD Poynting flux in
    the corona, signature of a ``hot'' jet.}
    \label{bernoulli1}
\end{figure}
\begin{figure}[t]
  \resizebox{\hsize}{!}{\includegraphics[angle=-90]{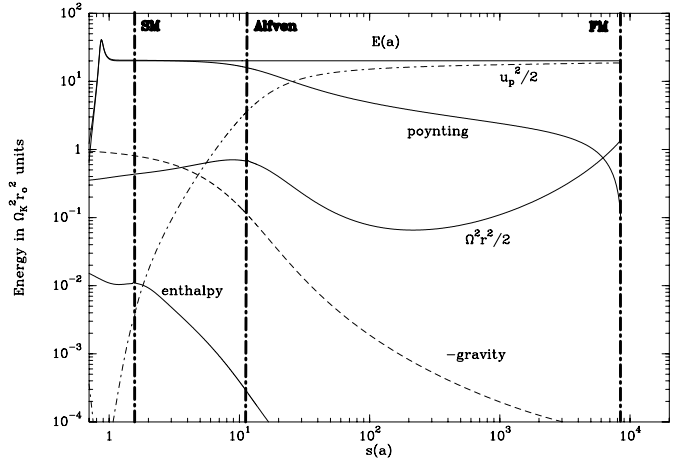}}
  \caption{Same plot than in Fig.~(\ref{bernoulli1}), but for the coldest jet
    ($f=0.002$) of Fig.~(\ref{Evo}). The enthalpy does not play any role in
    the jet energetics. This is the signature of a ``cold'' jet.}
    \label{bernoulli2}
\end{figure}
\begin{figure}
    \includegraphics[angle=-90,width=\columnwidth]{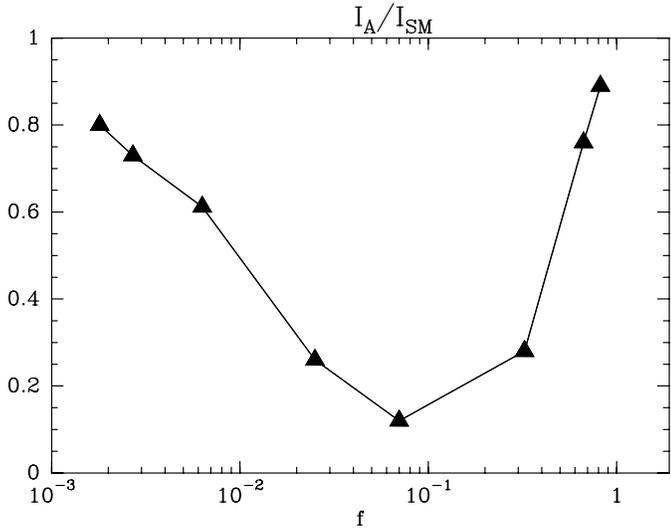}
    \caption{Ratio $I_A/I_{SM}$ of the current at the Alfv\'en point to
      the current at the slow-magnetosonic point as a function of the
      entropy parameter $f$, for the same MAES as in Fig.~(\ref{Evo}). As
      $f$ increases, there is a transition from one regime to the other.
      The first regime corresponds to ``cold'' jets, where the thermal
      energy is insignificant and the mass load is small. The second regime
      corresponds to ``hot'' jets, where a large thermal energy supplants
      the Poynting flux in propelling matter in the sub-Alfv\'enic
      region.} 
      \label{I1}
\end{figure}

\subsection{The self-similar ansatz}

\begin{figure}[t]
  \resizebox{\hsize}{!}{\includegraphics[angle=0]{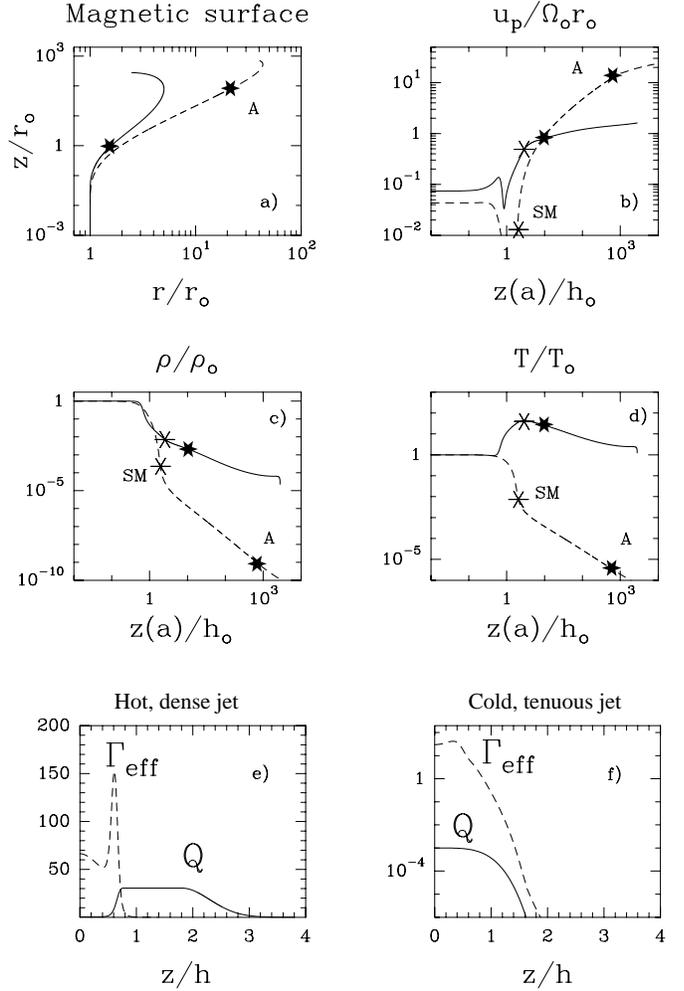}}
  \caption{Plots of various characteristics of the two extreme (``hot'' and
    ``cold'') jets presented in Sect.~4.3: {\bf a} poloidal cross-section
    of the magnetic surfaces anchored at a radius $r_o$; {\bf b} poloidal
    velocity of the plasma along a magnetic surface in units of the
    Keplerian speed at its footpoint; {\bf c} and {\bf d}, plasma density
    and temperature along a magnetic surface normalized to their values at
    the disc midplane. The ``hot'', dense jet is drawn in solid line while
    the ``cold'', tenuous jet is in dashed line. The cross symbolizes the
    locus of the slow-magnetosonic (SM) point and the star the locus of the
    Alfv\'en (A) point. Pannels {\bf e} and {\bf f} show the effective
    (viscous and Ohmic) heating term $\Gamma_{eff}$ and the prescribed
    entropy source $Q$ at a constant radius, normalized to the same
    fiducial quantity.}
    \label{partie}
\end{figure}

In this section, we briefly describe the radial self-similarity that
enables us to consider both the resistive, viscous disc and the ideal 
MHD jet, including an exact treatment of all dynamical terms. Any physical
quantity $A(r,z)$ is written as  
\begin{equation}
A(r,z) = A_e\left(\frac{r}{r_e}\right)^{\alpha_A}F_{A}(x)
\end{equation}
where $x = z/h(r)$ and $r_e$ is the outer radius of the magnetized
disc. This special ansatz allows to separate the variables so that the full
MHD equations become a set of non-ordinary (they contain three
singularities) differential equations. The values of the radial exponents
(given in Ferreira \& Pelletier \cite{FP93a}) are all linked to the
ejection index $\xi$. For example, the magnetic flux becomes
\begin{equation}
a(r,z) = a_e\left(\frac{r}{r_e}\right)^{\beta}F_a(x)
\end{equation}
with $\beta= 3/4 + \xi/2$. Inside this framework, the entropy source $Q$
must be written as
\begin{equation} 
Q = Q_{e}\left(\frac{r}{r_e}\right)^{\xi - 4}F_{Q}(x)
\end{equation}
\noindent where, using Eq.~(\ref{equi}), its value at the disc midplane
must satisfy
\begin{eqnarray}
Q_o & =& Q(r,z=0) = \left ( \frac{\gamma}{\gamma -1} + \alpha_P \right )
\frac{P_o u_{po}}{r} \nonumber \\
&= & \left ( \frac{\gamma}{\gamma -1} +\xi - \frac{5}{2} \right )
\frac{P_o u_{po}}{r} \ . 
\end{eqnarray}
This is a very strong constraint which also occurs in self-similar ADAF
models. There $\xi=0$, so an ADAF must have $\gamma \neq 5/3$ in order to
get some advected heating at the disc midplane. Here, we choose an
adiabatic index $\gamma = 5/3$ and scale the vertical profile $F_Q(x)$
with the parameter $f$. We display in Fig~.(\ref{Q}) a typical profile
corresponding to a discrepancy between cooling and heating processes,
occuring mainly at the disc surface. We did however play with slightly
different vertical profiles (see Fig.~(\ref{partie})).

The set of MHD equations has three critical points in the ideal MHD
regime. These critical points are the slow-magneto\-sonic, the Alfv\'en and
the fast-magnetosonic points. The solutions presented here share the
same behavior than in previous studies (Ferreira \cite{F97}, CF). The
magnetized flow passes through the slow and Alfv\'en points (the regularity
conditions impose the values of $\mu$ and $\xi$) and then encounters the
fast-magnetosonic point in a recollimation motion. None of these solutions
pass through the last critical point (see Vlahakis et al, in preparation).

\subsection{From magnetically-driven to thermally-driven jets}

As showed in Sect~3.1, the entropy generation increases the plasma pressure
gradient, thereby allowing an enhanced vertical mass flux. In particular,
jets are now possible in configurations where the magnetic pinching force
would be overwhelming and squeeze the disc. Thus, thermal effects
significantly enlarge the parameter space of MAES ($\varepsilon$,
$\alpha_m$, ${\cal P}_m$, $\chi_m$). 

In order to illustrate such an effect (strong magnetic compression balanced
by entropy generation), we choose a MAES configuration described by the
following parameters: $\varepsilon=0.1$, $\alpha_m=1.5$ ($\alpha_m^*=0.8$),
${\cal P}_m=1$, $\chi_m=1.5$. The chosen disc aspect ratio  is
relevant in the astrophysical systems we consider (see Appendix
C). Figure (\ref{Evo}) shows relevant quantities obtained with the same
vertical profile $F_Q$ (see Fig.~(\ref{Q})) and increasing values for $f$. 

First, a minimum heating ($f \sim 0.001$) is here required otherwise the
thermal plasma gradient cannot counteract the magnetic pinching force. Note
that this minimum heating decreases with decreasing $\alpha_m$ (because of
the weaker toroidal magnetic compression). Second, as expected, both the
mass flux $\kappa$ and coronal temperature $T^+$ grow as $f$
increases. Their maximum values depend here on the chosen set of disc
parameters. In a consistent way, the magnetic lever arm $\lambda$
diminishes with $f$. Third, as $f$ increases, thermal effects become
significant and there is a transition between ``cold'' and ``hot''
outflows. Indeed, when $f$ is very small, the jets show the usual
characteristics of a ``cold'' outflow, namely a fast magnetic rotator
($\omega_A >1$), a large magnetic lever arm $\lambda$, a small mass load
$\kappa$ and a coronal temperature $T^+$ smaller than that in the disc. At
the opposite, when $f$ is close to unity, the jets are slow magnetic
rotators ($\omega_A < 1$), have small magnetic lever arms, high mass loads
and high coronal temperatures.

We will label as ``hot'', jets whose initial enthalpy plays a significant role
in their launching. We display in Figures (\ref{bernoulli1}) and
(\ref{bernoulli2}), the Bernoulli constants $E(a)$ and their different
components, for the two extreme jets obtained in Fig.~(\ref{Evo})
(respectively $f=0.82$ and $f=0.002$). In the ``hot'' jet, the enthalpy at
the base is of the same order than the MHD Poynting flux. The jet
acceleration in the sub-Alfv\'enic region is mostly due to the decrease of
the enthalpy. This shows the importance of thermal effects, although the 
Blandford \& Payne criterion (opening angle larger than $30^o$) is, here,
still verified. In the ``cold'' jet however, the enthalpy is negligeable 
and the acceleration is due to the decrease of the poloidal current. 

In order to clearly distinguish ``cold'' (magnetically-driven) from ``hot''
(thermally-driven) jets, we look at the decrease along the jet of the
poloidal current, defined as 
\be
I = \frac{2 \pi}{\mu_o} r B_{\phi}  \ .
\ee
Indeed, this current is a measure of the transfert of MHD Poynting flux
into kinetic energy flux (see Ferreira \cite{F97}). We plot in
Fig.~(\ref{I1}) the ratio $I_A/I_{SM}$ of the current still available at 
the Alfv\'en point to the current provided at the base of the flow (at the
slow-magnetosonic point). For small $f$ (here $f <0.1$), this ratio remains
high because a tiny fraction of mass has been loaded in the jet. 
As $f$ increases, this ratio decreases because the enhanced mass load
requires more magnetic energy to be used. But at some point ($f >0.1$), 
the thermal energy reservoir supplants the MHD Poynting flux in propelling
matter until the Alfv\'en point. Thus, the ratio is now increasing because
of this new energy supply. The two regimes shown in Fig.~(\ref{I1}) define
a ``cold jet'' and a ``hot jet'' zone.

\begin{figure}[ht]
   \resizebox{\hsize}{!}{\includegraphics[angle=-90]{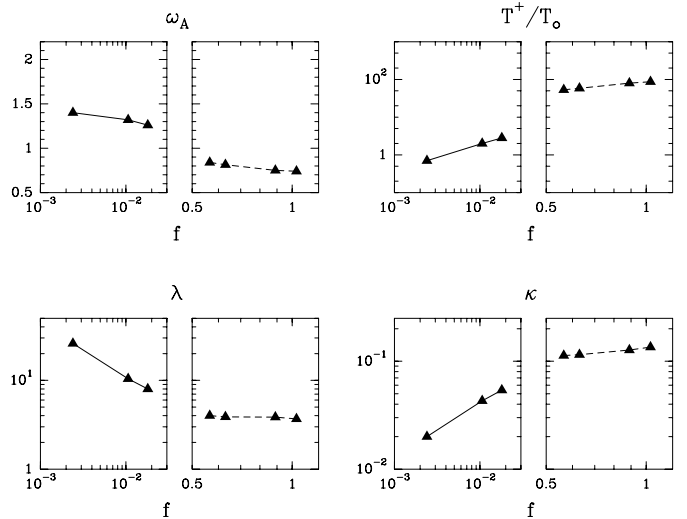}}
   \caption{Same plots than in Fig.~(\ref{Evo}), but for a weaker 
     turbulence level $\alpha_m= 1.2$. The two jet regimes are now
     splitted. ``Cold'' jets exist only for $f<0.02$, ``hot'' jets
     only for a large entropy generation, $f > 0.5$.} 
   \label{EvoII}
\end{figure}
\begin{figure}[t]
   \resizebox{\hsize}{!}{\includegraphics[angle=-90]{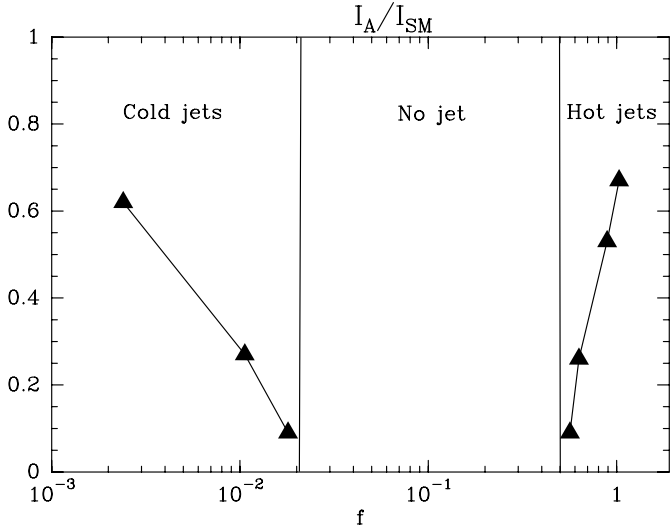}}
   \caption{Same plot than in Fig.~(\ref{I1}), but for a MAES with a weaker
     turbulence level $\alpha_m= 1.2$. A forbidden zone appears where
     both the MHD Poynting flux and the thermal energy are too small to
     propel the mass loaded.}  
   \label{I2}
\end{figure}

For a given MAES, there is always an upper limit to the entropy generation
(besides the overall energy conservation constraint) coming from the
requirement that $g_A$ must always remain smaller than unity (ie. $B_{\phi}$
remains negative). Indeed, the available current at the Alfv\'en point
writes 
\be
\frac{I_A}{I_{SM}} = g_A \frac{\lambda}{\lambda -1} \ .
\ee
The increase of this ratio for increasing $f$ is mainly due to the increase
of $g_A$ (see Fig.~(\ref{Evo})). Thus, there is a maximum $f$ allowed
(which can be smaller than unity) for steady-state MAES.

A transition from ``cold'' to ``hot'' jets requires to cross a zone
where a large mass load (provided by the enhanced plasma pressure gradient)
is still accelerated. This is only achieved if a sufficient amount of
toroidal magnetic field is present ($B_{\phi}$ is the most relevant
ingredient of the Poynting flux) and/or if the thermal energy reservoir is 
enough. However, depending on the magnetic configuration, the jet mass load
could increase with $f$ faster than the enthalpy. As a result,
configurations with a weak toroidal magnetic field would not be able to
steadily launch jets in a zone where the enthalpy is not sufficient. As an
example, we provide in Figures (\ref{EvoII}) and (\ref{I2}) a magnetic
configuration with a smaller turbulence level ($\alpha_m=1.2$), hence a
toroidal magnetic field smaller than in the previous case. This time,
no solution has been found between $ 0.02 < f < 0.5$.

\subsection{Two extreme MAES}

Because of the extreme sensibility of the mass load $\kappa$ to the entropy
source $Q$, we will not give a parameter space including these thermal
effects. Indeed, both the amplitude $f$ and the vertical profile $F_Q$
should be obtained by the resolution of a realistic energy equation
including cooling and heating terms. These terms (especially the cooling
function) depend critically on the astrophysical context studied. Our
purpose here is to show the broad range of available solutions by providing 
two extreme MAES configurations.

\subsubsection{A dense, slow ``hot'' jet}

In this extreme solution, a huge ($f=1$) entropy generation occurs above
the disc surface, see pannel e) in Fig.~(\ref{partie}). This might be
achieved with some sort of non-local coronal heating, efficiently
transporting upwards the thermal energy released in the underlying layers
of the disc.  

For this pedagogical example, we choose a magnetic configuration
described by the following disc parameters: $\varepsilon=0.1$, $\alpha_m=4$
($\alpha_m^*= 1.8$), ${\cal P}_m=1$, $\chi_m= 1.4$. It exhibits straight
poloidal magnetic field lines inside the disc and a factor $q=1.012$, leading
to a torque ratio $\Lambda= 2.6$ (comparable magnetic and viscous
torques). The required magnetic field must be quite small, $\mu= 0.26$
(ie. a plasma beta of 7.7 at the disc midplane).

The enormous release of thermal energy allowed a fantastic ejection, namely
$\xi= 0.456$. The corresponding jet parameters are a mass load $\kappa=
1.04$, a magnetic lever arm $\lambda= 1.9$ and a rotator parameter
$\omega_A= 1.65$. Although this solution is ``hot'', it is still a fast
rotator. This is because the toroidal field at the disc surface is high, $|
B_{\phi}^+ | = 0.94 B_o$, despite straight poloidal field lines (large
$\alpha_m$, see CF).   

\begin{figure*}[ht]
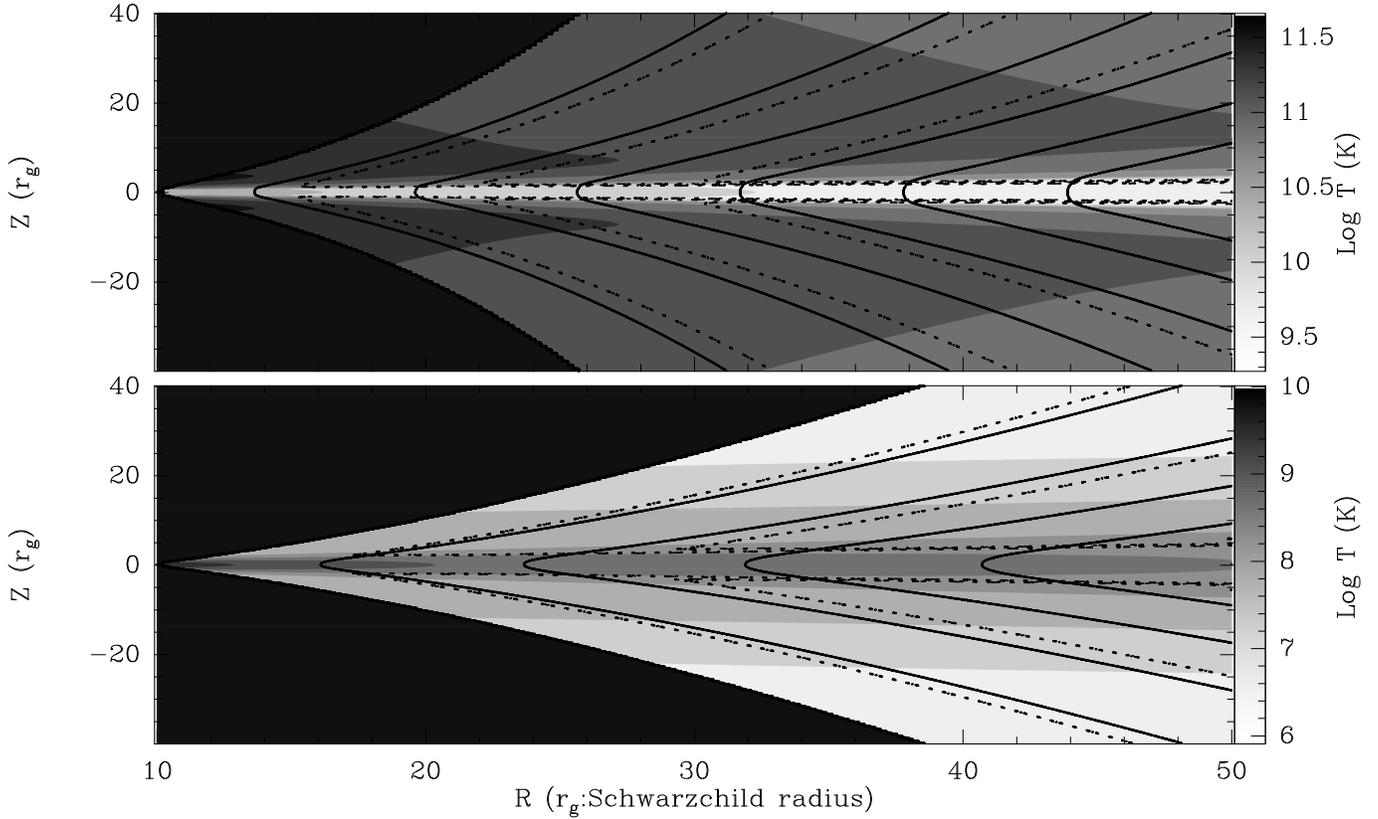

   \resizebox{\hsize}{!}{\includegraphics[angle=-90]{relat1.epsi}}
   \resizebox{\hsize}{!}{\includegraphics[angle=-90]{relat2.epsi}}    
   \caption{Cross-sections of two extreme magnetized accretion-ejection
     structures (MAES), settled around a neutron star of $M_*= 
     1.4M_{\odot}$ and fed with an accretion rate of $\dot{M}_{ae}=
     10^{-12}\ M_{\odot}/yr$. The poloidal magnetic field is shown in solid
     lines, streamlines are dashed and the grey contours represent the
     logarithm of the plasma temperature. Both MAES have the same
     temperature at the disc midplane. The upper image is an extremely
     ``hot'' MAES ($f=1$) with an entropy source located above the disc
     surface. This solution has a huge ejection efficiency of $\xi=0.456$,
     plasma reaching an asymptotic velocity 1.6 times the Keplerian speed
     $\Omega_o r_o$ at the magnetic field lines footpoint $r_o$. The lower
     image is a ``cold'' MAES ($f=5\ 10^{-5}$) with no heating source above
     the disc. Such a solution has $\xi=0.001$ with asymptotic velocities
     23 times $\Omega_o r_o$.}
   \label{relat}
\end{figure*}

The disc corona is hot and quite dense and the jet does not widen much
before recollimation occurs (see Fig.~(\ref{partie})). This widening is
however sufficient to produce jets where most of the available energy has
been transferred into poloidal kinetic energy, namely
\be
u_{p_\infty} \simeq \sqrt{2 E(a)}
\label{vterm}
\ee
along every magnetic surface. The presence of a large thermal energy allows
a terminal velocity larger than $\Omega_o r_o\sqrt{2\lambda - 3}$, which is
the characteristic value obtained in ``cold'' flows ($\Omega_o r_o$ is roughly
the Keplerian speed at the footpoint $r_o$ of the magnetic surface).
Figure (\ref{relat}) shows a cross section of such a solution, settled
around a compact object (upper image). Even in this extremely ``hot''
solution, the jet opening angle at the transition between resistive and
ideal MHD regimes is $\theta^+ = 37 ^o$, thus still fulfilling the
Blandford \& Payne condition for ``cold'' jets. 

The global energy budget is here exceedingly simple. Since $f=1$, all the
accretion power is finally converted into jet kinetic power, $P_{acc} = 2
P_{jet} \simeq 0.58\ P_{lib}$. Again, we do not claim this is a realistic
solution, but we want to point out at the extreme sensibility of the produced
jets to the disc energetics.

\subsubsection{A tenuous, fast ``cold'' jet}

Here, we show an example of a jet launched from a MAES with an almost
negligible entropy generation, namely $f=5\ 10^{-5}$ see pannel f) in
Fig.~(\ref{partie}). This slight discrepancy between heating and cooling
processes occurs only inside the disc. Thus, this solutions requires no
energy source above the disc, neither a coronal heating nor an
illumination. 

The MAES is described by the following disc parameters: $\varepsilon=0.1$,
$\alpha_m=0.3$ ($\alpha_m^*= 0.2$), ${\cal P}_m=1$, $\chi_m= 0.047$. It
exhibits a very strong curvature of the poloidal magnetic field lines
inside the disc and a factor $q=0.25$, leading to a torque ratio $\Lambda=
15.67$ (dominant magnetic torque at the disc midplane). Here, the required
magnetic field is closer to equipartition, $\mu=0.745$ (ie. a plasma beta of
2.7 at the disc midplane).

In this case, the strong compression due to the vertical magnetic pressure
gradient almost forbids the formation of a jet. The presence of a slight
entropy generation however allows an ejection with an efficiency $\xi=
0.001$. The corresponding jet parameters are a tiny mass load $\kappa= 2.8
\ 10^{-4}$, a huge magnetic lever arm $\lambda= 440$ and a rotator 
parameter $\omega_A= 1.2$. Again, this is not a ``very'' fast rotator
despite the small mass load, because of the small value of the toroidal
field at the disc surface ($| B_{\phi}^+ | = 0.12 B_o$).

The jet is ``cold'', with a very strong adiabatic cooling due to its large
widening (see Fig.~(\ref{partie})). This widening resulted in a very
efficient acceleration of the ejected plasma, leading to an asymptotic
poloidal velocity $\sim 23$ times the Keplerian speed at the field lines
footpoint. This is because of the huge magnetic lever arm $\lambda$. In
particular, relativistic speeds could be expected if such a MAES is settled
at the vicinity of a compact object. Indeed, if we define $\sigma_\infty$
as the asymptotic ratio of the MHD Poynting flux to the kinetic energy
flux, one gets an asymptotic Lorentz factor of 
\be
\gamma_{\infty}= 1 + \frac{2 \lambda -3}{12(1 + \sigma_\infty)} \left (
  \frac{3 r_g}{r_o} \right ) \ ,
\ee
where $r_g=2GM_*/c^2$ is the Schwarzschild radius of the central
object. Of course, a full calculation of the relativistic MHD equations
should be undergone in order to get the value of $\sigma_\infty$. As an
example, if some equipartition is achieved ($\sigma_\infty=1$, Li et
al. \cite{Li92}), one could in principle get jets with asymptotic Lorentz
factors ranging from $\sim 36$ ($r_o \sim 3 r_g$) to $\sim 4$ ($r_o \sim 30
r_g$).  

Figure (\ref{relat}) shows a cross section of such a solution, settled
around a compact object (lower image). The opening angle $\theta^+$
is quite large, $53^o$, typical of a ``cold'' outflow. The global energy
budget is here more subtle than in the previous case. Indeed, we obtain an
accretion power of $P_{acc} \simeq 1.09  \ P_{lib}$, jets fed with a total
power of only $2 P_{jet} \simeq 0.45 \ P_{lib}$ with the remaining power
being radiated at the disc surfaces, namely $2 P_{rad} \simeq 0.64
\ P_{lib}$. Thus, we do have a dissipative MAES producing ``cold''
jets. The very strong curvature of the poloidal magnetic field lines yields
a strong toroidal current, hence a large Ohmic dissipation. Moreover, the
large anisotropy of the MHD turbulence ($\chi_m=0.047$), provides also
a large dissipation of the toroidal magnetic field (see
Eq.~(\ref{gammaeff})). Hence, the dissipated power $P_{diss}$, which is here
converted into radiation, is quite large. Equation (\ref{jetpower}) shows
that the exact amount of energy powering the jets depends on the ratio
$| B_{\phi}^+/q B_o |$. Here this ratio is equal to $0.48$, namely a
generated toroidal magnetic field at the disc surface half the expected
value (measured by $q$). This decrease of the toroidal magnetic field is
due to this enhanced magnetic diffusivity $\nu_m'$.

\section{Conclusion}

In this paper, we removed a basic assumption of most theoretical
works dealing with axisymmetric, stationary Keplerian discs launching
jets (MAES). Indeed, the disc energy equation was usually neglected, using 
either adiabatic (Casse \& Ferreira \cite{Fab}) or isothermal (Wardle \&
K\"onigl \cite{Ward}, Li \cite{LiI}, \cite{LiII}, Ferreira \cite{F97})
magnetic surfaces. 

We used a self-similar ansatz allowing us to take into account all
dynamical terms in the resistive, viscous disc as well as in the ideal MHD
jet. The transition between the accretion and the ejection motions is
self-consistently described. We modelled the entropy increase along each
streamline by prescribing the vertical profile of the heat effectively
converted into entropy. This is an approach very similar to that used in
ADAF models, with a significant distinction. Indeed, in the case of MAES,
the main part of the heat released in the disc could be absorbed by the
jet instead of being advected onto the central object.

The possibility to steadily produce jets lies in the rather subtle disc
vertical equilibrium. Both gravity and the chosen bipolar magnetic topology
vertically pinch the accretion disc. Thus, mass can only be lifted from the
disc by the plasma pressure gradient. Because a local entropy generation
enhances this gradient (see Eq.~(\ref{equi})), the amount of mass escaping
the disc can be much larger. Alternatively, tiny mass loads, impossible to
achieve in previous MAES calculations, are now reachable. Indeed, a huge
magnetic compression can be now balanced by some amount of
heating. Moreover, there is a second important consequence of entropy
generation, namely enthalpy generation. This enthalpy is carried away by
the outflowing material and can be converted afterwards into kinetic
energy. To summarize, entropy generation results in dynamical (measured by
the ejection index $\xi$) and thermal (measured by the coronal temperature
$T^+$) effects. For the first time, this is self-consistently illustrated
in a global energy budget of MAES. This paper has illustrated that
heating is as important in steady jet formation from a accretion disc as in
steady outflow release around embedded sources (e.g. Lery et al. \cite{Lery}).

We are now in a position to provide clear diagnostics on steady-state MHD
simulations of jets from Keplerian accretion discs. Indeed, all effects
acting on the ejection process itself are taken into account. Of course,
this study was done under the self-similar assumption, but the links
between MHD invariants are local and, as such, independent of this
ansatz. Besides, in a quasi-Keplerian disc where gravity (which is
self-similar) is the dominant driving term, we expect self-similar
solutions to be quite reasonable. For example, local jet parameters
(mass load $\kappa$, magnetic lever arm $\lambda$ and rotator parameter
$\omega_A$) as well as the shape of the Alfv\'en surface (ie. conical), are
comparable to stationary solutions obtained with full 2D, time-dependent
numerical simulations of ``cold'' jets (Krasnopolsky et al. \cite{Kras}). 

Ustyugova et al. (\cite{Uts}), using another MHD code and starting with a
split-monopole poloidal magnetic field, obtained a non-conical Alfv\'en
surface. However, their local jet parameters remain comparable to those
obtained in this study. Indeed, they have larger mass loads and smaller
magnetic lever arms, but with a prescribed temperature satisfying
$C_s^+/\Omega_{Ko}r_o = (T^+/T_o)^{1/2}\varepsilon \geq 0.3$. Because the
authors assumed a flat disc (no flaring at all, hence $\varepsilon \ll 1$),
this boundary condition requires a coronal temperature larger than the
temperature at the disc midplane. As a consequence, a significant entropy
generation must take place in the underlying accretion disc.  

Ouyed \& Pudritz (\cite{Ouyed}) also obtained stationary jets from
Keplerian discs with full 2D, time-dependent MHD simulations. They assumed 
power-law magnetic fields for their boundary conditions at the jet basis
and ``cold'' adiabatic magnetic surfaces. Surprisingly, they obtained jets
with very large mass loads ($\kappa> 0.6$) and magnetic lever arms very
small ($\lambda < 3.5$). This large mass flux is compatible with a large
toroidal magnetic field at the jet basis ($\sim 1.5$ times the vertical
field). However, the disc vertical equilibrium cannot handle such a strong
pinching in a non-dissipative disc ($f \ll 1$, see CF). Besides, we showed
in Sect.~3.3 that thin discs with $T^+ \leq T_o$ require $\lambda > 4$, a
value smaller than those obtained in these simulations. One possible way to
reconcile our analytical result with their numerical simulation is to
realize that the jet basis used in all MHD simulations of that kind is {\bf
  not} the disc surface but some surface above it. In particular, it could
be located above a region where a coronal heating (allowing a large
mass load $\kappa$) was followed by a very strong cooling (negligible
enthalpy at the ``jet basis''). One must then be very cautious when
identifying boundary conditions as being the disc surface. Indeed, the real
issue of jet formation is the mass load provided by the underlying
accretion disc. 
 
The effects of entropy generation, theoratically shown in this paper, were
numerically confirmed by the self-similar solutions. For example, the
surplus of thermal energy, obtained with a significant parameter $f$,
allowed jets that would be labelled as slow magnetic rotators ($\omega_A <
1$). We did not intent to explore any parameter space because of the
extreme sensibility of the mass load $\kappa$ to the entropy source $Q$
(both vertical profile and amplitude, measured by $f$). For example, a very
weak entropy generation triggered inside the disc could give birth to very
tenuous, fast jets while a huge entropy generation just above the disc
surface could make dense, slow jets. 

Without exploring the parameter space, we nevertheless obtained a much wider
range of MAES parameters than under an adiabatic assumption (CF), namely
mass loads $10^{-4} \leq \kappa \leq 1$ (ie. $0.001 \leq \xi < 0.5$),
magnetic lever arms $1.9 \leq \lambda \leq 440$ and magnetic rotator
parameters $0.7 < \omega_A < 2.2$. Such huge ejection efficiencies could 
have a strong impact on disc spectral energy distributions. Indeed, if the
disc is optically thick, the effective temperature writes
(Ferreira \& Pelletier \cite{FP95}) 
\be
T_{eff} \propto r^{-\frac{3}{4} + \frac{\xi}{4}} \ .
\ee  
Thus MAES with large ejection efficiencies would display a flat
distribution (possibly up to $T_{eff} \propto r^{-1/2}$). However, only a
precise treatment of the plasma energetics (including all heating, cooling
and transport processes) provides the exact value of the disc ejection
efficiency $\xi$. This can only be done for a given astrophysical 
context, since all these energetic processes (and possibly MHD turbulence
$\alpha_m$, ${\cal P}_m$ and $\chi_m$) strongly depend on the 
plasma properties. Therefore, any quantitative modelling of a MAES must
include a complete treatment of the energy equation.

\begin{acknowledgements}
The authors would like to thank Guy Pelletier and Jean Heyvaerts for
useful comments and fruitful discussions.
\end{acknowledgements}

\appendix
\section{Global energy conservation}

\begin{figure}[t]
  \includegraphics[angle=0,width=\columnwidth]{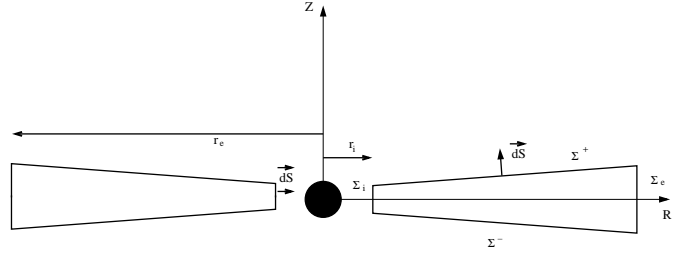}
  \caption{Cross section of the volume $V$ occupied by the disc and the
    corresponding bounding surfaces (lateral $\Sigma_i$ and $\Sigma_e$, upper
    $\Sigma^+$ and lower $\Sigma^-$). The upper and lower surfaces 
    correspond to the location of a vanishing radial velocity (typically
    at $z^+ \sim h(r)$).}
  \label{contour}
\end{figure}

The local energy conservation equation can be written as
\begin{eqnarray} 
\vec{\nabla} \cdot (\rho\vec{u_p}(\frac{u^2}{2}\ +&\Phi_G&+\ H)\ +\ 
 \vec S_{MHD} \ -\  \vec u \cdot \mathsf{T})
= \nonumber\\
& &\rho  T\vec{u_p} \cdot \vec{\nabla} S\  -\  \Gamma_{eff} \ .
\label{Econs}
\end{eqnarray}
where $\vec S_{MHD}$ is the MHD Poynting flux, $\mathsf{T}$ the turbulent
stress tensor, $H$ the enthalpy, $S$ the specific entropy and
$\Gamma_{eff}$ the effective Ohmic and viscous heating (see Sect.~3.1). The
second law of thermodynamics (\ref{madaf}),  
\begin{eqnarray}
\rho T\vec{u_p} \cdot\vec{\nabla} S &=& \Gamma_{eff}\ +\
(\Gamma_{turb}\ -\ \Lambda_{turb})\ +\ \Gamma_{ext} \nonumber \\
& & -\  \vec{ \nabla} \cdot \vec S_{rad}
\end{eqnarray}
is then incorporated into Eq.~(\ref{Econs}). To get the global energy
conservation, we integrate this equation on the whole volume $V$ occupied
by the disc (Ferreira \cite{F94}, Henriksen \cite{Dick}). We thus define
$\Sigma^+$ and $\Sigma^-$ as the disc surfaces 
and $\Sigma_i$ ($\Sigma_e$) the lateral surfaces at $r=r_i$ ($r=r_e$, see
Fig.~(\ref{contour})). The disc surface is here precisely defined at the
locus where the radial velocity vanishes, namely where $u_r(r,z^+= x^+h(r))
=0$. 

After integration, we get 
\be
P_{acc} + P_{ext} + P_{turb} = 2P_{jet} + 2P_{rad}
\ee
\noindent where the accretion power $P_{acc}$ is the power liberated by the
accretion flow (the difference between what comes in at $r_e$ and goes out
at $r_i$), $P_{ext}= \int_V \Gamma_{ext} dV$ is the power provided by an
external source, $P_{turb}= \int_V (\Gamma_{turb} - \Lambda_{turb})dV$ is a
source of turbulent energy, $P_{jet}$ is the power carried by the
outflowing matter and $P_{rad}= \int_{\Sigma^{\pm}} \vec{S}_{rad} \cdot
\vec{dS}$ is the luminosity emitted at the disc surfaces.  

The available turbulent power $P_{turb}$ comes from the small scale
turbulent motions, but is carried along by the laminar accretion
flow. Therefore, there is also a flux of specific turbulent energy
${\cal E}_{turb}$ crossing the inner and outer sides of the disc, which
implies 
\be
P_{turb} = - \int_{\Sigma_i} \rho\vec{u_p} {\cal E}_{turb} \cdot \vec{dS}
- \int_{\Sigma_e} \rho\vec{u_p} {\cal E}_{turb} \cdot \vec{dS}
\ee
We redefine the accretion power $P_{acc}$ by including $P_{turb}$
inside. The total accretion power is then 
\begin{eqnarray}
  P_{acc} & = & - \int_{\Sigma_i} \left [ \rho\vec{u_p} {\cal E} + \vec
    S_{MHD} + \vec S_{rad} - \vec u \cdot \mathsf{T} \right ] \cdot \vec{dS}
  \nonumber\\
  & & - \int_{\Sigma_e} \left [ \rho\vec{u_p} {\cal E} + \vec
    S_{MHD} + \vec S_{rad} - \vec{u} \cdot \mathsf{T} \right ] \cdot \vec{dS}
\end{eqnarray}
where the expression of the MHD Poynting flux in the resistive regime is
given in Ferreira \& Pelletier (\cite{FP93a}) and
\be
{\cal E}= \frac{u^2}{2} + \Phi_G + H + {\cal E}_{turb}
\ee
is the advected specific energy of the flow (laminar and turbulent). The
jet power is given by 
\begin{eqnarray}
  P_{jet} & = & \int_{\Sigma^{\pm}} \left [ \vec{S}_{MHD} + \rho
    \vec{u_p}{\cal E}  \right ] \cdot \vec{dS} = \int_{\Sigma^{\pm}} \rho
    \vec{u_p}E(a) \cdot \vec{dS} \nonumber \\
& &  
\end{eqnarray}
where $E(a)$ is the Bernoulli invariant (\ref{Bernou}) since, at the disc
surface, the flow is in ideal MHD regime. Finally, we obtain  
\be
P_{acc} + P_{ext} = 2P_{jet} + 2P_{rad}
\label{cons}
\ee
which is the global energy conservation equation of a MAES. 

\section{Self-similar energy conservation}

In the self-similar framework used here, the global energy conservation can
be expressed in terms of the fiducial power 
\be
P_{lib}= \eta_{lib}\frac{GM_*\dot{M}_{ae}}{2 r_i}\ \ \  \mbox {where}\ \ \  \eta_{lib}=
\frac{\left(\frac{r_i}{r_e}\right)^\xi - \left(\frac{r_i}{r_e}\right)}{1-\xi}
\ee
\noindent and local disc parameters. The accretion power is a sum of
various terms, namely $P_{acc}= P^{mec}_{acc} + P^{th}_{acc} +
P^{MHD}_{acc} + P^{rad}_{acc} + P^{vis}_{acc} + P_{turb}$. The mechanical
contribution to $P_{acc}$ is
\be
\frac{ P^{mec}_{acc}}{P_{lib}}  \simeq   (1 - \xi)(2 - \delta^2 -
m^2_s\varepsilon^2) 
\ee
\noindent $\delta=\Omega_o/\Omega_K$ ($\Omega_K$ is the Keplerian rotation
rate) and $m_s=u_{po}/\Omega_Kh_o$ being defined in CF. The advection of
enthalpy by the accreting material decreases the liberated energy by a
factor 
\be
\frac{P^{th}_{acc}}{P_{lib}} \simeq - 2(1-\xi) \frac{\gamma}{\gamma-1}
\varepsilon^2  
\ee
while the MHD Poynting flux gives a positive contribution
\be
\frac{P^{MHD}_{acc}}{P_{lib}} \simeq \frac{1-\xi}{2}\frac{\Lambda}{1+\Lambda}
\left | \frac{B_{\phi}^+}{q B_o} \right | \tan\theta^+ \delta^2\varepsilon 
\ee  
where $\theta^+$ is the angle between the magnetic field lines and the
vertical axis at the disc surface ($x=x^+$). The presence of a turbulent
viscosity produces a radial angular momentum flux through the disc, hence a
flux of energy. This contribution to $P_{acc}$ is 
\be
P^{vis}_{acc} = - \frac{P_{lib}}{1+\Lambda} \frac{3\delta^2}{2 \eta_{lib}} 
\frac{r_i}{r_e}  - \int_{\Sigma_i} (\vec{u}\cdot \mathsf{T}) \cdot \vec{dS} 
\ee
\noindent where the second term is evaluated at the inner edge of the
disc. There, it is usually assumed that the viscous torque is equal to zero,
$\mathsf{T}(r_i)=0$, leaving only a loss of energy at the disc outer edge.

Finally, the total power carried by the jets is
\begin{eqnarray}
\frac{2P_{jet}}{P_{lib}} & = & \frac{\Lambda}{1+\Lambda} \left |
  \frac{B^+_{\phi}}{qB_o}\right| (1 -\varepsilon x^+ \tan\theta^+) \delta^2
\nonumber\\
& & +\  \frac{2\gamma}{\gamma-1} \frac{T^+}{T_o}\xi\varepsilon^2 \nonumber\\
& & \ -\ \xi \left [ \frac{2}{(1 + \varepsilon^2x^{+2})^{1/2}}\right. 
\left. - \ \left( \frac{u^+}{\Omega_o r_o}\right)^2 \right]
\end{eqnarray}
\noindent where $u$ is the total velocity of matter.

\section{Disc thickness}

In this appendix, we  focus on the calculation of the disc aspect ratio
imposed by the energy  transport. To illustrate the full range of possible
astrophysical contexts, we first consider systems where the disc plasma is
``cold'', dense and slow (young stellar objects) then systems where the
disc plasma is ``hot'', tenuous and fast (around compact objects). 

It is useful to identify which heating process dominates. In
the case of MAES, we are able to estimate the ratio of viscous heating to
ohmic heating at the disc midplane, namely 
\be
\left .\frac{\eta_v\left|r \vec{\nabla}\Omega\right|^2}{\eta_m
    J_{\phi}}\right|_{z=0}  = 
\frac{9}{4{\cal P}_m\mu(1+\Lambda)^2\varepsilon^2}
\label{C1}
\ee
\noindent and
\be
\left.\frac{\eta_v\left|r\vec{\nabla}\Omega\right|^2}{\eta'_m J_{p}^2}\right|_{z=0} = \frac{3\mu}{\varepsilon^2} \ .
\label{C2}
\ee
For thin ($\varepsilon \ll 1$) or even slim disc
($\varepsilon < 0.3$), we have ${\cal P}_m\sim 1$, $\mu \sim 1$, $\chi_m
\leq 1$ and $\Lambda \leq \varepsilon^{-1}$. The viscous heating is always
 dominant for solutions with comparable torques ($\Lambda\sim 1$).

 We have the same approach than in the standard model
(Shakura \& Sunyaev \cite{Shak}) since the viscous heating is
the dominant heating source (we neglect ohmic heating) and since the disc is
assumed to be optically thick. Although the energy transport processes are
unknown, we can safely assume that the fraction $(1-f)$ of heating is fully
transported to the disc surface where it is radiated at the photosphere, 
\be
\sigma T^4_{eff} = \int_{o}^{h}\eta_v\left(r\frac{d\Omega}{dr}\right)^2dz \
.
\ee
The effective temperature $T_{eff}$ is related to
the central temperature via the optical depth of the disc
($T^4_{eff}\sim T^4/\tau$). Because we have assumed that the total pressure
gradient equals the plasma pressure gradient, there is a direct link
between the central temperature $T_o$ and the ``photospheric'' height of the
disc. 

We thus validate the assumption
of thin discs in the inner regions and checked the assumption of an
optically thick disc.

\subsection{Young Stellar Objects}  

In this context, the opacity of the disc can be fitted by a
Rosseland grain opacity such as $\kappa_R=0.1T^{1/2}$ $cm^2.g^{-1}$ (see
Bell \& Lin\cite{Bell}). We obtain 
\begin{eqnarray}
& &\varepsilon_{phot} \equiv \frac{h_{phot}}{r}  =\nonumber \\
0.07(1-f)^{1/9}& \alpha_v^{-1/9}&
\left(\frac{\dot{M}_{a}}{10^{-7}M_{\odot}/yr}\right)^{2/9} 
\left(\frac{M_*}{M_{\odot}}\right)^{-1/3}.      
\end{eqnarray} 
\noindent The resulting optical depth is 
\begin{eqnarray}
\tau &=& \int_{o}^{h_{phot}}\rho\kappa_R dz \simeq \kappa_R \rho_o h_{phot} \nonumber\\ 
  = 
 & &\frac{0.925}{\alpha_v^{10/9}(1-f)^{1/9}(1 +
 \Lambda)}\left(\frac{\dot{M}_{a}}{10^{-7}M_{\odot}/yr}\right)\nonumber \\
& &\times\left(\frac{M_*}{M_{\odot}}\right)^{1/3}\left(\frac{r}{1 AU}\right)^{-1} 
\end{eqnarray} 
This disc is indeed optically thick if the parameter $\alpha_v$ and/or
the torque ratio $\Lambda$ are small ($\alpha_v\sim 10^{-2}$ in YSO).
   
\subsection{Compact objects (AGN, microquasars)}

In the inner region of such systems, the Thomson opacity dominates. We thus
obtain
\begin{eqnarray}
\varepsilon_{phot} &=& \frac{0.6}{\alpha_v^{1/10}}(1-f)^{1/10} \nonumber\\
& \times&\left(\frac{\dot{M}_{a}}{M_{\odot}/yr}\right)^{1/5}
\left(\frac{M_*}{M_{\odot}}\right)^{-3/10}
\left(\frac{r}{3r_g}\right)^{1/20} 
\end{eqnarray}
\noindent where $r_g=2GM_*/c^2$ is the Schwarzchild radius. If we
take $M_*= 10^7M_{\odot}$ and $\dot{M}_{a}=1 M_{\odot}/yr$ as fiducial
values for AGN, we get $\varepsilon_{phot} \simeq 3\ 10^{-3}(1-f)^{1/10}\alpha_v^{-1/10}$.  

\noindent In the case of microquasars, we assume $M_*= 10 M_{\odot}$ and
$\dot{M}_{a}= 10^{-7} 
M_{\odot}/yr$, so we get  $\varepsilon_{phot}
\simeq  10^{-2}(1-f)^{1/10}\alpha_v^{-1/10}$. 

The corresponding optical depth is
\begin{eqnarray}
\tau = \frac{2.4\
  10^{8}}
  {\alpha_v^{4/5}(1-f)^{1/5}(1+\Lambda)}&&\left(\frac{r}{3r_g}\right)^{-3/5}\left(\frac{\dot{M}_a}{M_{\odot}/yr}\right)^{3/5}\nonumber \\
& &\times\left(\frac{M_*}{M_{\odot}}\right)^{-2/5}\ .
\end{eqnarray}
As long as the photospheric disc aspect ratio remains smaller than unity, these
discs remain optically thick. Anyway we must be aware that the photospheric
height given in this appendix is a lower limit since we have neglected 
other heating sources. Indeed  we can
see, by looking at Eq.(\ref{C1}) and (\ref{C2}), that MAES dominated by the
magnetic torque ($\Lambda \geq \varepsilon^{-1}$) can provide Ohmic
heating larger (but of the same order) than viscous heating. We thus choose
a disc aspect ratio larger than those calculated here, namely
$\varepsilon=0.1$ (a value at the intersection of thin and slim discs).

\end{document}